\documentclass[aps,prb,showpacs,groupedaddress,superscriptaddress,twocolumn]{revtex4}
\usepackage{amsfonts}
\usepackage{graphicx}
\usepackage{latexsym}
\usepackage{makeidx} 
\usepackage{amsmath} 
\usepackage{amssymb}
\usepackage{graphicx}
\usepackage{color}
\begin{document}

\title{Generation method of a photonic NOON state with quantum dots in coupled nanocavities}
\author{Kenji Kamide}
\email{kamide@iis.u-tokyo.ac.jp}
\author{Yasutomo Ota}
\affiliation{%
 Institute for Nano Quantum Information Electronics (NanoQuine), University of Tokyo, Tokyo 153-8505, Japan}
\author{Satoshi Iwamoto}
\author{Yasuhiko Arakawa}%
%\homepage{http://wwwacty.phys.sci.osaka-u.ac.jp/~ogawa/cover.html}
\affiliation{%
 Institute for Nano Quantum Information Electronics (NanoQuine), University of Tokyo, Tokyo 153-8505, Japan}
\affiliation{%
 Institute of Industrial Science, University of Tokyo, Tokyo 153-8505, Japan
}%
%\affiliation{
\date{\today}
% It is always \today, today,              %  but any date may be explicitly specified  
\begin{abstract}
We propose a method to generate path-entangled $N00N$-state photons from quantum dots (QDs) and coupled nanocavities.
In the systems we considered, cavity mode frequencies are tuned close to the biexciton two-photon resonance. Under appropriate conditions, the system can have the target $N00N$ state in the energy eigenstate, as a consequence of destructive quantum interference. The $N00N$ state can be generated by the resonant laser excitation.
This method, first introduced for two-photon $N00N$ state ($N=2$), can be extended toward higher $N00N$ state ($N>2$) based on our recipe, which is applied to the case of $N=4$ as an example. 
\end{abstract} 
\maketitle 
\noindent 
\section{\label{sec1} Introduction } 
Fundamental question to the coherence of laser light~\cite{Glauber} developed quantum optics~\cite{Mandel}---a research field on quantum light---which has explored functionality and application of light inaccessible by classical light.
Single photons are indispensable for quantum information processing~\cite{KLM, KoKrev, BosonSampling} and quantum communication~\cite{BB84, BB84-exp, Waks, Miyazawa}. 
Multi-photon source allows for multi-photon imaging for medical purpose, making possible imaging deep inside human brain with both increased penetration length and reduced damaging tissue~\cite{Denk, Horton}. 
Recent observation showing sensitivity of biological photoreceptors to photon statistics~\cite{Krivitsky1, Krivitsky2, Smart}, indicates potential impact of using quantum light in research of biology, as recognized in terms of quantum biology~\cite{Ball}. 
The situation in turn is accelerating theoretical studies for new quantum light source~\cite{Munoz} and new applications~\cite{Carreno}. 

An attractive application that takes the advantage of quantum light is the one using $N00N$ state.
Photonic $N00N$ state~\cite{Dowling} is a kind of entangled Fock (number) states of two orthogonal modes, defined by
\begin{eqnarray}
|N00N \rangle \equiv \frac{ |N,0 \rangle +|0, N \rangle}{\sqrt{2}}= \frac{(a_1^\dagger)^N +(a_2^\dagger)^N}{\sqrt{2 \times (n!)}} |{\rm vac} \rangle,
\end{eqnarray} 
where $a_1^\dagger$ and $a_2^\dagger $ are the creation operators of mode 1 and 2, and $|{\rm vac} \rangle$ is the vacuum state.
In particular, path-entangled $N00N$ state, in which the two modes are located in different optical paths, can be used for phase-supersensitive quantum lithography~\cite{Boto} and quantum metrology~\cite{O’Brien, Giovannetti}, as shown by the entanglement-enhanced microscope~\cite{Ono}.
By using $N00N$-state photon source in interferometry, the phase error can be reduced to so-called Heisenberg limit that cannot be achieved by classical laser light~\cite{Kok-Heisenberg, Nagata}.
Since the phase sensitivity increases with the photon number, such application requires generation of $N00N$ states with large $N$.
However, realization of an efficient source of $N00N$ state with large $N$ has been a challenging issue, especially in optical regime (in contrast in microwave regime~\cite{Wang, Su}).

Popular approach to generate photonic $N00N$ state is based on the use of photons generated from spontaneous parametric down conversion (SPDC) processes in nonlinear $\chi_2$-crystal~\cite{Kwiat-SPDC}, linear optical elements, and postselection~\cite{Kok, Afek, Nagata}.  
However, the approach results in the low generation rate and the generation occurs in non-deterministic way.
This is essentially because the SPDC photon source is not true quantum light source~\cite{Mandel}.

Another approach is to use true quantum light source based on quantum emitters with strong optical nonlinearity.
Quantum dots (QDs) are ideal solid-state quantum emitters~\cite{Buckley}, whose emission rate can be increased further by embedding them inside photonic nanocavities~\cite{Purcell, Englund, Strauf, Nomura1, Kamide1}. 
Efficient generation method of polarization entangled two-photon $N00N$ state ($N=2$) was proposed theoretically by using two polarization modes of nanocavity~\cite{Gies, Valle3}.
On the other hand, method using QD-nanocavity systems for higher $N00N$ states with $N>2$ has not been reported.

In this paper we propose a method to generate path-entangled $N00N$ states with QDs in coupled nanocavities~\cite{Bayer, Reithmaier, Vasconcellos, Ishii, Majumdar1, Sato, Majumdar2, Bose}, called photonic molecules~\cite{Bayer}.
In our method, photons emitted from each of the two cavities, which can be guided into two separated paths, can form the path-entangled $N00N$ state.
Key idea in this method is utilizing the quantum interference between multiple quantum paths, being similar to concept used for pure single-photon generation in coupled-cavity systems with weak optical nonlinearity~\cite{Liew, Bamba}.
This method has some advantages over the past approaches; the high-rate and on-demand emission of two-photon $N00N$ state becomes possible. Moreover, extension toward higher-$N00N$-state generation is possible for general case of $N (>2)$.  

This paper is organized as follows. 
In Sec.~\ref{sec2}, we firstly show our method for $2002$-state generator in system with a QD in coupled nanocavities.
Then, we evaluate performance of the $2002$-state generator by simulating purity and available generation rate.
In Sec.~\ref{sec3}, we generalize the proposed method to $N00N$ state with $N>2$, where simulation of $4004$-state generator is also presented as example.
This paper is summarized and conclusion is made in Sec.~\ref{sec4}, where we also mention the comparison with existing method on $2002$-state generator~\cite{Gies, Valle3} and future issue.

\section{\label{sec2}Generation method of 2002 state}
Here we show how to generate two-photon $N00N$ states in QD-coupled cavity systems. 
The generation method is explained in several steps. 
In the following four subsections, we explain our method by showing how to prepare 2002-state generating state ``2002-GES'' (named later) as an energy eigenstate of the system, how to excite 2002-GES, decay dynamics of the excited 2002-GES, and the available detection rate of the 2002-state photons emitted out from the cavities,  in Sec.~\ref{2002design}, \ref{2002exc}, \ref{2002decay}, and \ref{2002rate}, respectively.
Through the discussion, we show the essence of this scheme, which can be generalized to the case of $N>2$ in Sec.~\ref{sec3}.
  
\subsection{\label{2002design}Preparation of 2002-GES in QD-coupled-cavity systems}
\begin{figure}[tb]
\begin{center}  
\includegraphics[scale=0.3]{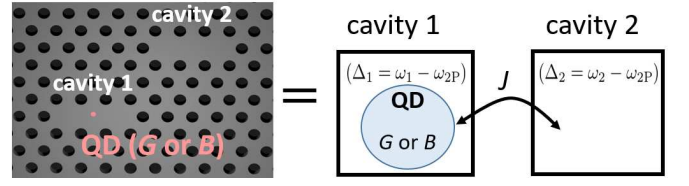} {} 
\end{center}
\vspace{-2 mm}
\caption{\label{fig1} Schematic of the QD-coupled nanocavity system for two-photon $N00N$ state (right), which can be realized in PhC platform (left)}
\end{figure}

System we consider is composed of two nanocavities, cavity 1 and cavity 2, coupling through tunneling (tunneling rate $J$), and a QD in cavity 1 (coupling constant $g$), as shown in Fig. \ref{fig1}. 
This system can be realized in nanocavities, using micropillars~\cite{Bayer, Reithmaier, Vasconcellos}, microdiscs~\cite{Ishii}, and photonic crystals (PhCs)~\cite{Majumdar1, Majumdar2, Sato, Bose}.
Cavity resonance frequencies ($\omega_{1}$ and $\omega_{2}$ for cavity 1 and cavity 2, respectively) are tuned close to the QD-biexciton two-photon resonance, $\Delta_{1(2)} \equiv \omega_{1(2)} -\omega_{\rm 2P} \sim 0$~\cite{Ota} where $\omega_{\rm 2P} \sim (2 \omega_X-\chi)/2$, and $\hbar \omega_X$ and $\hbar \chi$ are a single-exciton energy and biexciton-binding energy (we set $\hbar =1$ hereafter). 
In this case, by truncating the irrelevant single-exciton states, the effective Hamiltonian is approximated by~\cite{Valle1}, 
\begin{eqnarray}
H_{\rm eff} &=& \sum_{j=1,2} \Delta_{j} a^\dagger_j a_j +J(a^\dagger_1 a_2+a^\dagger_2 a_1)\nonumber \\
&& +g_{\rm 2P} \left( a_1^2 |B \rangle \langle G | +   |G \rangle \langle B| (a_1^\dagger)^2 \right) \label{Heff2}
\end{eqnarray}
where $g_{\rm 2P}=4g^2/\chi$, $a_j$ is an annihilation operator of photons in cavity $j$($=1,2$), and $|G \rangle$ and $|B \rangle$ represent the vacuum and biexciton states of the QD. 
This approximation can be used if cavities are tuned to the biexciton two-photon resonance, as far as $g \ll \chi/2$, whose validity is confirmed theoretically~\cite{Valle1} and experimentally~\cite{Ota}.

Typical value of the coupling constant is $g \sim 100$ $\mu$eV~\cite{Kuruma} in PhC platform, and the biexciton binding energy $\chi$ ranges between sub to few meV~\cite{Ota} (which can be electrically controllable~\cite{Trotta}). 
For $g=100$ $\mu$eV and $\chi=0.8$ meV, the two-photon coupling constant $g_{\rm 2P}$ is estimated to be 50 $\mu$eV.
Strong two-photon nonlinearity is observable if the cavity loss rate $\kappa$ is smaller than $g_{\rm 2P}$.
$g_{\rm 2P}$ and $1/g_{\rm 2P}$ define characteristic energy and time scale for this system.
The cavity detuning $\Delta_j$ can be controlled in some ways, e.g. by temperature tuning technique~\cite{Vuckovic1, Englund} and by xenon gas deposition technique~\cite{Mosor}. 
Tunneling parameter $J$ depends on a distance between cavities, meV order for direct coupling~\cite{Majumdar1} and tens of $\mu$eV for waveguide mediated coupling~\cite{Sato}.
Especially for the latter case, $J$ is electrically controllable~\cite{Konoike1} with high precision in a range of $\mu$eV, using an extra control cavity~\cite{Konoike2}.

Our strategy for generating pure $2002$ state is to find conditions for three parameters, $\Delta_1/g_{\rm 2P}$, $\Delta_2/g_{\rm 2P}$, and $J/g_{\rm 2P}$, so that the two-photon $N00N$ state can be included in the eigenstate of $H_{\rm eff}$ in Eq.~(\ref{Heff2}).
If we could find the condition, the eigenstate, which emits $2002$-state photons out from the cavities, can be exclusively excited in the system by resonant pumping in presence of the strong two-photon nonlinearity ($\kappa < g_{\rm 2P}$).
In this sense, we shall call such eigenstate ``$2002$-state generating eigenstate (2002-GES)''.

Noticing that $H_{\rm eff}$ commutes with the total excitation number operator, $[N_{\rm tot}, H_{\rm eff} ]=0$ with $N_{\rm tot}=2 | B \rangle \langle B | +a_1^\dagger a_1+a_2^\dagger a_2$, we focus on the eigen equation in the Hilbert subspace of $N_{\rm tot}=2$, $H_{\rm 2P}|E \rangle=E|E \rangle$, where 
\begin{eqnarray}
H_{\rm 2P} &=&
\left(
    \begin{array}{cccc}
      0 & \sqrt{2} g_{\rm 2P} & 0 & 0 \\
     \sqrt{2} g_{\rm 2P} &2 \Delta_1 & \sqrt{2} J & 0\\
      0 & \sqrt{2} J & \Delta_1+\Delta_2 &\sqrt{2}  J \\
      0 & 0 & \sqrt{2} J & 2\Delta_2
    \end{array}
  \right)  ,
\end{eqnarray}
and $| E \rangle=(A_B, A_{20}, A_{11}, A_{02}) = A_B |B,00\rangle +A_{20} |G,20\rangle +A_{11} |G,11\rangle +A_{02} |G,02 \rangle$. 
A state vector, $|i,n_1 n_2 \rangle$, represents QD state $i$ ($=B$ or $G$) with $n_{1}$ and $n_2$ photons in cavity 1 and cavity 2, respectively.
The 2002-GES which we want to prepare is an eigenstate $| E \rangle$ with $|A_{20}|=|A_{02}|$ and 
\begin{eqnarray}
A_{11}=0. \label{Req1}
\end{eqnarray}
Under the requirement, Eq.~(\ref{Req1}), the third row of $H_{\rm 2P}|E \rangle=E|E \rangle$ gives $J(A_{20}+A_{02})=E A_{11}=0$.
Therefore $|A_{20}|=|A_{02}|$ is automatically fulfilled with Eq.~(\ref{Req1}) for any nonzero $J$.
There are two solutions to the eigen equation under Eq.~(\ref{Req1}), which are labeled by $s(=\pm)$.
For the detuning parameters satisfying 
\begin{eqnarray}
\Delta_2 &=& \frac{ \Delta_1 + s \sqrt{\Delta_1^2+2 g_{\rm 2P}^2 } }{2}, \label{Cond1}
\end{eqnarray} 
one of the four eigenstates of $H_{\rm 2P}$ and the eigenenergy are given by
\begin{eqnarray}
|E_s \rangle &=& \cos{\varphi_s} |B,00\rangle + \sin{\varphi_s} \left( \frac{|G,20\rangle-|G,02\rangle}{\sqrt{2}}\right), \quad  \label{eq:2002state} \\
E_s &=& \Delta_1 + s \sqrt{\Delta_1^2+2 g_{\rm 2P}^2 },
\end{eqnarray} 
where $\varphi_s \equiv \arctan (E_s/g_{\rm 2P})$.

\begin{figure}[tb] 
\begin{center}  
\includegraphics[scale=0.3]{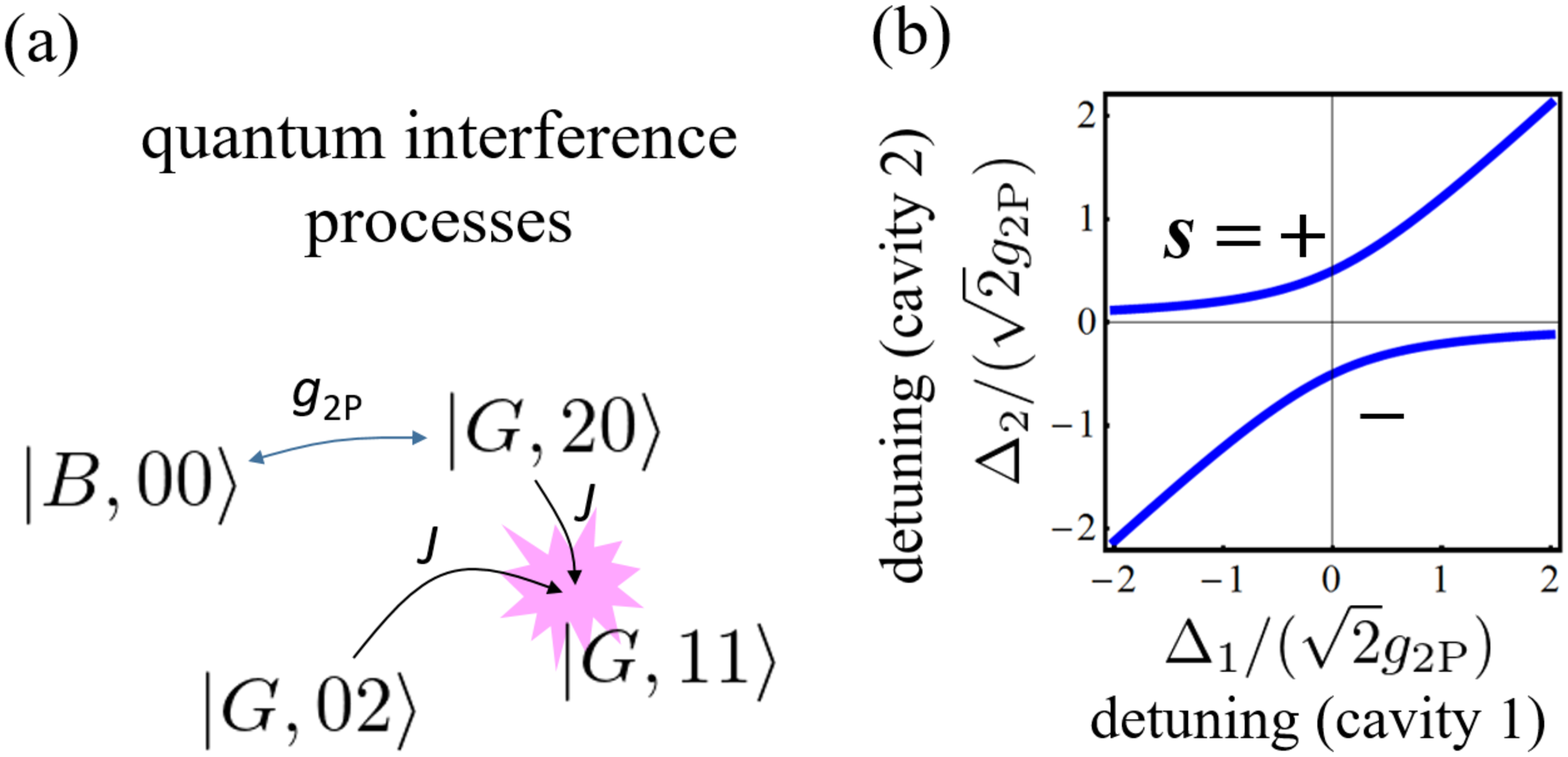} {} 
\end{center}
\vspace{-5 mm}
\caption{\label{fig2} (a) Quantum interference processes in the system described by Eq.~(\ref{Heff2}). 
Arrows indicate the direction of coherent population flow at an initial time starting from the prepared $2002$-GES ($| E \rangle$ with $A_{11}=0$).
In the figure, arrows connected to $|G,11 \rangle$ are unidirectional since the initial state does not contain $|G,11 \rangle$, hence with no outflow from it. 
 (b) The required condition for the cavity detunings, $\Delta_1$ and $\Delta_2$, so that the two-photon $N00N$ state become one of the four energy eigenstate.}
\end{figure} 

The condition and the solution can be interpreted as follows; 
The requirement on cavity tuning, Eq.~(\ref{Cond1}) (plotted in Fig.~\ref{fig2} (b)), arises in order to satisfy Eq.~(\ref{Req1}).
The 2002-GES, $| E_s \rangle$, is the superposition of the biexciton state, $|B,00 \rangle$, and photonic 2002 state, $\frac{|G,20\rangle-|G,02\rangle}{\sqrt{2}}$, with probability ratio $(\tan{\varphi_s})^2$, while only the latter component contributes to photon emission through the cavity loss.
It is remarkable that all the above condition, eigenstates, and eigenenergies are independent of $J$.
This originates from the fact that the requirement in Eq.~(\ref{Req1}) is fulfilled for any value of nonzero $J$ as far as $A_{20}=-A_{02}$. 
The third row of the Schr\"odinger equation $i\frac{d}{dt}|E \rangle=H_{\rm 2P}|E \rangle$, 
\begin{eqnarray}
i\frac{d}{dt} A_{11} = J(A_{20}+A_{02})
\end{eqnarray}
shows that if $A_{20}=-A_{02}$ is satisfied, the quantum interference between two processes, $|G,20\rangle \to |G,11\rangle$ and $|G,02\rangle \to |G,11\rangle$, becomes fully destructive and the generation rate of the unwanted state $|G,11\rangle$ vanishes irrespective of $J$ (Fig.~\ref{fig2} (a)).

While $E_s$ does not depend on $J$, energies of the other three two-photon eigenstates do.
Therefore, the 2002-GES can be excited exclusively by resonant laser pumping with a frequency $\omega_p-\omega_{\rm 2P}=E_s/2$, if $J$ is selected properly to isolate it from others (as shown below in Fig.~\ref{fig5} (a)) so that photon blockade effect~\cite{Birnbaum, Shamailov} can work.

\subsection{\label{2002exc}Excitation of 2002-GES in the system}
The 2002-GES can be excited resonantly thanks to the strong optical nonlinearity. 
This can be confirmed by solving numerically the quantum Master equation (QME) within Born-Markov approximation~\cite{Carmichael}, taking into account the cavity decay processes.
In the simulation, we have made several assumptions.
To simplify discussion, we assumed that the energy dissipation is dominated by cavity loss, while the spontaneous emission of the QD excitons and biexcitons directly to free space can be negligible in PhC platform, due to photonic band gap effect. 
In addition,  we consider a case where the two cavities have the same loss rate $\kappa$.
These assumptions do not change our main conclusion.

\begin{figure}[tb] 
\begin{center}  
\includegraphics[scale=0.33]{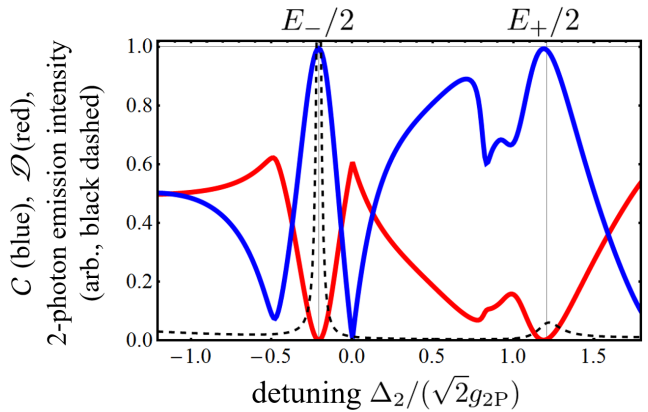} {}  
\end{center}
\vspace{-7 mm}
\caption{\label{fig3} Two-photon emission properties are shown as a function of $\Delta_2$ under weak cw pumping on cavity 2 with $\omega_p-\omega_{\rm 2P}=\Delta_2$: Concurrence $\mathcal C$ (blue solid), trace distance $\mathcal D$ (red solid), and normalized two-photon emission intensity (black dashed). Simulation is performed for parameters $ (\kappa,   J ,  \Delta_1, \Omega ) =(0.1, 2, 1, 0.05) \times \sqrt{2}g_{\rm 2P}$. }
\end{figure}

In a frame rotating with the excitation laser frequency $\omega_p$, the QME is given by
\begin{eqnarray}
\frac{\rm d }{{\rm d} t} \rho&=&  \mathcal{L} \rho = \frac{1}{i}[H'_{\rm eff}+H_{\rm pump},\rho]  +\mathcal{L}_{\rm loss} \rho,  \label{eq:QME} \\
H'_{\rm eff} &\equiv&  H_{\rm eff} -(\omega_p-\omega_{\rm 2P})N_{\rm tot}. \label{eq:Heff'} 
\end{eqnarray}
Here, $\mathcal{L}_{\rm loss} \left(\equiv \kappa(\mathcal{L}_{a_1}+\mathcal{L}_{a_2}) \right)$ represents the cavity loss, and $H_{\rm pump} \left( =\Omega (a_2+ a_2^\dagger ) \right) $ describes cw pumping on cavity 2 with the Rabi field amplitude $\Omega$ (see also appendix~\ref{appendix A}). 
For a parameter set $ (\kappa,  J ,   \Delta_1 ,  \Delta_2,  \omega_p-\omega_{\rm 2P}, \Omega ) =(0.1, 2, 1, -0.207, -0.207, 0.05) \times \sqrt{2}g_{\rm 2P}$ satisfying the condition, Eq.~(\ref{Cond1})  ($\Delta_2=E_-/2$ for $s=-$), the cw laser excitation can generate the 2002-GES, which emits 2002-state photons. The quantum state of the emitted photons can be observed by the state tomography. In this case, the two-photon density matrix to be observed,  $\rho_{\rm tomo}$ (defined in appendix~\ref{appendix B}), is found to be
\begin{eqnarray*}
\left(
  \begin{array}{ccc}
0.500 & -0.001 + 0.018 i & -0.495 - 0.049 i \\
-0.001 - 0.018 i & 
  0.002 & -0.002 + 0.018 i \\
-0.495 + 0.049 i & -0.002 - 0.018 i &   0.498
  \end{array}
  \right),
\end{eqnarray*}
which is very close to those for pure 2002 states.
The purity of the generated 2002 state is quantified by the trace distance from the pure 2002 state $\mathcal D$ and concurrence  $\mathcal C$ ($\mathcal D=0$ and $\mathcal C=1$ correspond to an ideal case where pure 2002 state is generated: see appendix~\ref{appendix B} for the details).
The above result gives small trace distance, $\mathcal D =0.0007$, and high concurrence, $\mathcal C =0.995$.

%The eq.~(\ref{eq:Tomography2P}) can be taken as true two-photon density matrix in the weak pump regime where the detection probability of three or more photons at a time can be negligible.

In Fig. \ref{fig3}, in order to see how effective is the parameter tuning at the condition (Eq.~(\ref{Cond1})) is, we plotted the two-photon emission properties as a function of the cavity frequency $\Delta_2$. 
Here we consider weak cw pumping on cavity 2 (with the laser frequency $\omega_p=\Delta_2$).
It is clear that the concurrence $\mathcal C$ approaches unity and the trace distance $\mathcal D$ approaches zero just at the two points, $\Delta_2 = E_-/2 \approx -0.207 \times \sqrt{2}g_{\rm 2P}$ and $E_+/2\approx 1.207\times \sqrt{2}g_{\rm 2P}$, which locate at the peaks of two-photon emission intensity as well. 
This shows that the 2002-GES can be excited exclusively, hence the pure 2002-state photons can be generated, if the cavity frequencies are tuned at the condition. 
High contrast in the emission intensity at two peaks comes from the difference in the generated population and also in the decay rate (or emissivity) of the target states, $|E_s \rangle$, which depend also on the details of the energy level structure of intermediate one-photon states (see appendix~\ref{appendix C}).
%We note that both measure, $\mathcal C$ and $\mathcal D$ can be used to evaluate the purity of generated $N00N$-state photons. 

\begin{figure}[tb] 
\begin{center}  
\includegraphics[scale=0.30]{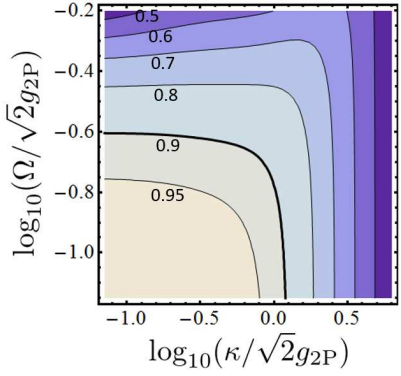} 
\end{center}
\vspace{-7 mm}
\caption{\label{fig4} Contour plot of the concurrence $\mathcal C$ as a function of cavity loss $\kappa$ and laser Rabi frequency $\Omega$, for the case of cw resonant pumping on cavity 2 ($\Delta_2=\omega_p-\omega_{\rm 2P}=E_-/2=-0.207 \times \sqrt{2}g_{\rm 2P}$). 
Simulation is performed for parameters $ (J ,  \Delta_1 ) =(2.0, 1.0) \times \sqrt{2}g_{\rm 2P}$. }
\end{figure} 
 
Figure \ref{fig4} shows the simulated concurrence $\mathcal C$ as a function of cavity loss $\kappa$ and the Rabi frequency $\Omega$, for $ (J,  \Delta_1 ) =(2.0, 1.0) \times \sqrt{2}g_{\rm 2P}$ which fulfill the condition,  Eq.~(\ref{Cond1}). 
High value of $\mathcal C$ is observed for small $\kappa$ and $\Omega$ ($\mathcal C>0.9$ for $\kappa/\sqrt{2}g_{\rm 2P}<1$ and $\Omega/\sqrt{2}g_{\rm 2P}<0.25$).
This can be understood as follows; the purity of the generated 2002-state photons becomes high for high-$Q$ cavities because the 2002-GES is energetically separated from other states and hence can be excited exclusively if the cavity linewidth $\kappa $ is smaller than the level spacings of $ \mathcal O ( J)$ and/or $\mathcal O ( g_{\rm 2P})$.
However, even with high-$Q$ cavity, the purity degrades with pumping strength due to a mixing of the higher-number Fock states for $\Omega>\mathcal O(J)$.

\begin{figure}[tb]  
\begin{center}  
\includegraphics[scale=0.42]{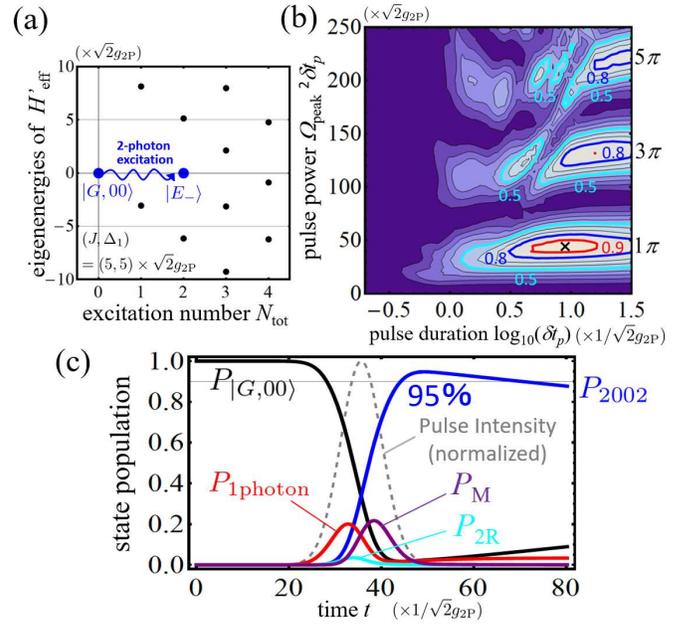} 
\end{center}
\vspace{-7 mm}
\caption{\label{fig5} (a) Set of eigenenergies and excitation number of the eigenstates of $H'_{\rm eff}$ in Eq.~(\ref{eq:Heff'}) for $ (J ,  \Delta_1, \Delta_2,\omega_p-\omega_{\rm 2P})=(5, 5,-0.05, -0.05)$. (b) Population of the 2002-GES ($P_{2002}$) generated just after the pulse excitation ($t=t_{\rm peak}+2 \delta t_p$), as a function of pulse power $\Omega_{\rm peak}^2 \delta t_p$ and pulse duration $\delta t_p$. (c) Rabi excitation dynamics ($\pi$-pulse) for $ (J ,  \Delta_1, \Delta_2 ,\omega_p-\omega_{\rm 2P}, \kappa,   ) =(5, 5,-0.05, -0.05, 0.1) \times \sqrt{2}g_{\rm 2P}$ with $\Omega_{\rm peak}^2 \delta t_p=45\times \sqrt{2}g_{\rm 2P}$ and  $\delta t_p =10^{0.95}/(\sqrt{2}g_{\rm 2P})$, corresponding to the parameters marked by a cross in (b). The half-width at half maximum of the Rabi pulse intensity is $(\ln 2/2)^{1/2} \delta t_p\approx 0.59 \delta t_p$. }
\end{figure} 

 As an alternative way, the 2002-GES can be excited in deterministic way by using short Rabi pulse, whereas the cw excitation discussed above is a probabilistic way.
Dynamics of the system during the pulsed generation is described by two-photon Rabi oscillation if one-photon excitation is negligible. This is possible with proper choice of $J$.
For example, for $H'_{\rm eff}$, there exist two one-photon eigenstates, 
\begin{eqnarray}
|{\rm 1P}, + \rangle &=& \cos{\phi} |G, 10 \rangle +\sin{\phi} |G, 01 \rangle, \label{1Pstate1} \\
|{\rm 1P}, - \rangle &=& -\sin{\phi} |G, 10 \rangle +\cos{\phi} |G, 01 \rangle, \label{1Pstate2}
\end{eqnarray}
with $\phi={\rm arctan} \left( \sqrt{1+(\frac{\Delta_1-\Delta_2}{2J})^2}- \frac{\Delta_1-\Delta_2}{2J}\right)$, 
whose eigenenergy is $E'_{{\rm 1P},\pm}=E_{{\rm 1P},\pm}-(\omega_p-\omega_{2P})$ with
\begin{eqnarray}
E_{{\rm 1P},\pm}=\frac{\Delta_1+\Delta_2}{2} \pm \sqrt{\left(\frac{\Delta_1-\Delta_2}{2}\right)^2+J^2}.
\end{eqnarray} 
As seen clearly, $E_{{\rm 1P},\pm}$ depends on $J$, while the energy of the 2002-GES, $E_s$, does not.
Therefore, if these one-photon states are detuned from the 2002-GES by tuning $J$, it is possible to excite the 2002-GES exclusively without generating one-photon states. 
Similarly, the 2002-GES can be detuned from the other two-photon eigenstates with proper choice of $J$.
As an example, we plotted in Fig.~\ref{fig5} (a) the eigenenergies of $H'_{\rm eff}$ in Eq.~(\ref{eq:Heff'}) sorted by the total number of excitations $N_{\rm tot}$, for $(J,\Delta_1)=(5, 5) \times \sqrt{2} g_{\rm 2P}$ and $\Delta_2=\omega_p-\omega_{\rm 2P}=E_-/2$. 
The resonance excitation from the ground state $|G,00 \rangle$ occurs to a state with eigenenergy of zero, whose candidate is only $| E_- \rangle$, the 2002-GES, for this parameter.

We perform a simulation on the pulsed Rabi dynamics by integrating numerically the Eq.~(\ref{eq:QME}), where the pulse shape is assumed to be a Gaussian function,  $\Omega(t)=\Omega_{\rm peak} \exp \left( -(t-t_{\rm peak})^2/\delta t_p^2 \right)$. 
In Fig.~\ref{fig5} (b), we plotted population of the 2002-GES generated just after the pulsed excitation, $\left. P_{2002} \right|_{t=t_{\rm peak}+2\delta t_p} \equiv {\rm Tr} \left(| E_- \rangle \langle E_-| \times \rho(t_{\rm peak}+2\delta t_p)\right) $, as a function of the pulse power $(\propto \Omega_{\rm peak}^2\delta t_p)$ and the duration $\delta t_p$.
We found several maxima located with almost equal intervals in the pulse power, with each corresponding to $1\pi$, $3\pi$, and $5\pi$ pulse conditions from the bottom of the plot.
The highest probability exceeds 90 percent with $\pi$-pulse excitation. 
In order to have high probability, the pulse intensity should not be too strong (to avoid the higher-number state mixing), and the pulse duration should be shorter than the state decay time, $\delta t_p < 1/\Gamma_{2002}$ ($\Gamma_{2002}$ will be given in Sec.~\ref{2002decay}), and longer than the tunneling time, $1/J <\delta t_p $. 
The last requirement comes from the fact that a short pulse duration results in the frequency broadening $1/\delta t_p$, degrading the success probability of the selective excitation of the targeted 2002-GES if the broadened linewidth exceeds the energy separation from other states, $\mathcal{O} (J)$. In this way, there exists an optimal pulse duration.
Fig.~\ref{fig5} (c) shows the dynamics of the state population with the optimal $\pi$-pulse excitation (corresponding to a cross in Fig.~\ref{fig5} (b)).
In this case, population of the 2002-GES, $P_{\rm 2002}$, reaches 95 percent and population of the other two and higher number states, $P_{\rm 2R}$  and $P_{\rm M}$, are suppressed to be less than 1 percent.
Demonstrating the deterministic generation with high-$Q$ nanocavities, especially with small $\kappa=0.1 \times \sqrt{2} g_{\rm 2P} $ for Fig.~\ref{fig5}(c) (i.e. $Q \approx 0.18 \times 10^6$ for $\omega_1 (\approx \omega_2)=1.3$ eV and $g_{\rm 2P}=50 \ \mu$eV), is challenging in current technology but will be reached in future~\cite{Takamiya, Ota2}. 

To summarize this section, the 2002-GES can be exclusively excited in the system in both non-deterministic and deterministic ways by the resonant laser excitation, if it is prepared as an eigenstate by appropriate cavity tuning (Eq.~(\ref{Cond1})).

%During a chirped pulse excitation, in the rotating frame oscillating with the frequency $\omega_p(t)$, the eigenstate (quasi-energy levels) of $H_{\rm eff}+H_{\rm pump}$ smoothly changes with time.
%Depending on the pulse parameters, the pulse intensity, duration, and chirp rate $\dot{\omega}_p$, the smooth change accompanies a transition from initial state to finial state after a pulse.
%It is possible to select the chirp rate $\dot{\omega}_p$, intensity, and duration, so that the dynamics is a adiabatic process while the cavity decay process is slow enough to be negligible within a pulse duration.
%The transition dynamics through such adiabatic process is called ``a rapid adiabatic passage''~\cite{Eastham}.  
%This method, applied successfully to determinisctic preparation of QD exciton states~\cite{ARPexp}, is also applicable to prepare the designed $N00N$ state here.  
%This can be done by a chirped laser pulse (i.e. $\Omega$ and $\omega_p$ have time dependence), through an adiabatic rapid passage~\cite{Eastham}. Fig. 4 (a) shows the simulated quantum dynamics for a Gaussian shape pulse with a half-width $T_p=*$ and a constant chirp rate $\alpha_p \equiv {\rm d}\omega_p/{\rm d}t=*$. 
%High probability generation (above 90  percent just after the pulse) of the 2002-GES. 

\subsection{\label{2002decay} Decay of 2002-GES via cavity leakage}
Once created by $\pi$-pulse excitation, population of the 2002-GES, $P_{2002}\equiv {\rm Tr} \left(| E_s \rangle \langle E_s|  \rho \right)$, gradually decreases due to free decay of photons from the cavity. 
Here we study in details the decay dynamics, which enables the evaluation of available rate of simultaneous two-photon detection (see Sec.~\ref{2002rate}). 

The free decay dynamics is described by a closed set of rate equations for $P_{2002}$, $P_{{\rm 1P},\pm}\equiv {\rm Tr} \left(| {\rm 1P},\pm \rangle \langle {\rm 1P},\pm |  \rho \right)$, and $P_{G,00}\equiv {\rm Tr} \left(| G,00 \rangle \langle G,00 |  \rho \right)$, which is derived from Eq.~(\ref{eq:QME}) by neglecting population of the other states with two or more photons ($P_{M}=P_{\rm 2R}=0$):
\begin{eqnarray}
\frac{\rm d}{{\rm d}t}P_{2002}&=&-\Gamma_{2002} P_{2002}  \\
\frac{\rm d}{{\rm d}t}P_{{\rm 1P},\pm}&=& (\Gamma_{2002}/2)P_{2002} -\Gamma_{{\rm 1P},\pm} P_{{\rm 1P},\pm} \\
\frac{\rm d}{{\rm d}t}P_{G,00} &=& \Gamma_{{\rm 1P},+} P_{{\rm 1P},+}+\Gamma_{{\rm 1P},-} P_{{\rm 1P},-},
\end{eqnarray}
where the decay rate $\Gamma_{2002}=2(\sin{\varphi_s})^2 \kappa$, and $\Gamma_{{\rm 1P},\pm}=\kappa$.
As it is clear from the expression of $\Gamma_{2002}$, decay rate of the 2002-GES is reduced from the bare rate of two-photon states $2\kappa$, by a factor $(\sin{\varphi_s})^2$ which is a fraction of the photonic component in $| E_s \rangle$ (whereas the biexcitonic fraction is $(\cos{\varphi_s})^2$, see Eq.~(\ref{eq:2002state})). 
In the case of the deterministic $\pi$-pulse geneneration, by solving the equations with $P_{\rm 2002}=1$ and $P_{{\rm 1P},\pm}=P_{G,00}=0$ at $t=0$, we found
\begin{eqnarray}
\rho (t) &=&P_{\rm 2002} (t)| E_s \rangle \langle E_s |
+\sum_{\sigma=\pm} P_{{\rm 1P},\sigma } (t)| {\rm 1P}, \sigma \rangle \langle {\rm 1P}, \sigma |  \nonumber \\
&&+P_{G,00 }(t)| G,00 \rangle \langle G,00 |, \label{sol:PIpulse}
\end{eqnarray}
where
\begin{eqnarray}
P_{2002}(t)&=&e^{-\Gamma_{2002} t},  \label{sol:P2002}  \\
P_{{\rm 1P},\pm}(t)&=& \frac{\Gamma_{2002} \left( e^{-\Gamma_{{\rm 1P},\pm} t}-e^{-\Gamma_{2002} t}\right) }{2(\Gamma_{2002}- \Gamma_{{\rm 1P},\pm} )} ,  \label{sol:P1} \\
P_{G,00}(t) &=& 1-P_{2002}-P_{{\rm 1P},+}-P_{{\rm 1P},-}. \label{sol:G}
\end{eqnarray}
\begin{figure}[t]  
\begin{center}  
\includegraphics[scale=0.28]{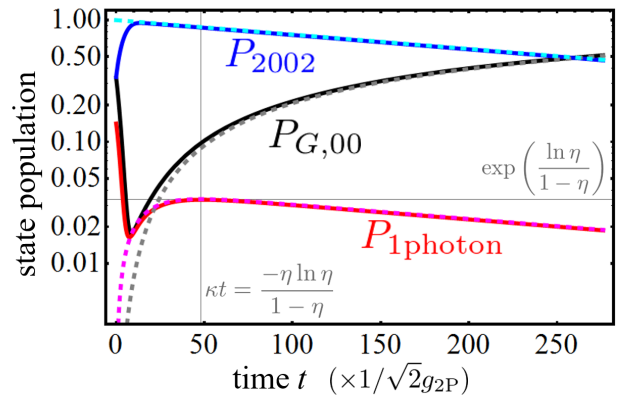} 
\end{center}
\vspace{-7 mm}
\caption{\label{fig6} Decay dynamics of the state population (long-time scale view of Fig.~\ref{fig5} (c)). 
We shift the origin of time ($t$=0) to the the pulse peak time ($t_{\rm peak}$ in Fig.~\ref{fig5} (c)). 
Solid and dashed lines show the numerical and analytic results (Eqs.~(\ref{sol:P2002})-(\ref{sol:G})), respectively. From analytic solution, the maximum one-photon population $P_{\rm 1photon}$ is given by $\exp \left(\frac{ \ln \eta}{1-\eta} \right) \approx 0.034$ (gray line), where $\eta=\Gamma_{2002}/\Gamma_{{\rm 1P},\pm}=2(\sin{\varphi_s})^2\approx 0.038 $ in this plot.}
\end{figure} 

Long time evolution of the population of the 2002-GES, $P_{2002}$, vacuum state, $P_{G,00}$, and one-photon state, $P_{\rm 1photon} (\equiv P_{{\rm 1P},+}+P_{{\rm 1P},-})$, are shown in Fig.~\ref{fig6}, where exact numerical results (solid lines) are compared with analytic results in Eqs.~(\ref{sol:P2002})-(\ref{sol:G}). 
It clearly shows that the long-time scale dynamics is well approximated by the analytic result (for $t>2 \delta t_p  =17.8 \times 1/\sqrt{2}g_{\rm 2P}$ when the Rabi excitation dynamics is negligible).
One photon population $P_{\rm 1photon}$ initially increases and monotonically decreases with time after a peak time.
By using Eq.~(\ref{sol:P1}) and $\eta \equiv\Gamma_{2002}/\Gamma_{{\rm 1P},\pm}=2(\sin{\varphi_s})^2$, the peak time and peak values are well approximated to be $\kappa^{-1} \frac{\eta \ln \eta}{1-\eta}$ and $ \exp \left(\frac{\ln \eta}{1-\eta} \right)$, respectively.
Thus, one-photon probability can be made small by tuning $\Delta_1$ so that $\eta$ be small. 
%Therefore, if necessary (which is not the case for application based on multi-photon interference, however), one-photon probability can be made small by tuning $J$ and $\Delta_1$ so that the photonic fraction of the 2002 state $(\sin{\varphi_s})^2(=\eta/2)$ is designed small since $\lim_{\eta \to 0}  \exp \left(\frac{\ln \eta}{1-\eta} \right)=0$.
%However, in that case, the emission rate $\Gamma_{2002}$ is also reduced.

\subsection{\label{2002rate}Two-photon simultaneous detection rate of emitted 2002-state photons}
Here we discuss available simultaneous detection rate of two photons emitted from the 2002-GES, whose measurements can be used in quantum state tomography (given by Eq.~(\ref{eq:Tomography2P}) in Appendix \ref{appendix B}) and also in applications of the phase-sensitive quantum metrology.
In experiments, rate of the simultaneous photodetection is measured as the number of multi-photon counting events within a small time window, $\Delta T_w$~\cite{Troiani, Valle3}.
While the detection rate increases with $\Delta T_w$, a visibility of the multi-photon interference will diminish due to an increase in the time uncertainty.
An important quantity we discuss here is the maximum detection rate, at which the real quantum state tomography shows the $N00N$-state correlation correctly with sufficient visibility. 

Under excitation of the 2002-GES with deterministic $\pi$-pulse, number of events for simultaneous detection of photons emitted from cavity $i$ and $j$ in time window $\Delta T_w$ is
\begin{eqnarray}
\mathcal{N}_{ij}&=&\int \int   \langle \hat{\mathcal{T}}_+ \hat{\mathcal{T}}_-  \hat{I}_i(t) \hat{I}_j(t')  \rangle \theta(\Delta T_w-|t-t'|) {\rm d}t {\rm d}t' \nonumber \\
&=&  \kappa^2  \int_0^{t_f} {\rm d}t \int_0^{\Delta T_w} {\rm d} \tau 
 \langle  a_i^\dagger (t) a_j^\dagger (t+\tau) a_j (t+\tau) a_i (t) \rangle \nonumber \\
&& + (i \leftrightarrow j), \nonumber \\
&=& \kappa^2  \int_0^{t_f} {\rm d}t \int_0^{\Delta T_w} {\rm d} \tau 
{\rm Tr}  \left( a_j^\dagger a_j e^{\left. {\mathcal L}\right|_{\Omega=0} \tau} [a_i \rho(t) a_i^\dagger    ]\right) \nonumber  \\
&& + (i \leftrightarrow j), \label{eq:Nij}
\end{eqnarray}  
where $t_f$ is an accumulation time of photo detection,  $\hat{I}_{i(j)}(t)=\kappa a^\dagger_{i(j)} (t) a_{i(j)} (t)$ is the Heisenberg operator of the emission rate from $i(j)$-th cavity at time $t$, $\theta(x)$ is Heaviside step function, $\hat{\mathcal{T}}_{+(-)}$ is a time ordering (anti-ordering) operation applied to the Heisenberg operators of $a (a^{\dagger})$, and $\langle \hat{X} \rangle \equiv {\rm Tr} (\hat{X} \rho_0)$ with an initially prepared 2002-GES, $\rho_0 = | E_s \rangle \langle E_s |$.
The time evolution of the density matrix, $\rho(t)$, is given by Eqs.~(\ref{sol:PIpulse})-(\ref{sol:G}).   
The accumulation time $t_f$ is longer than the decay time of the 2002-GES, $ t_f > 1/\Gamma_{2002}$, and is assumed equal to the pulse repetition time $\Delta T_{\rm rep}$ for repeated-pulse measurements. 
In this case, $t_f$ is replaced by $+ \infty$ in Eq.~(\ref{eq:Nij}).
For these assumptions, for diagonal part of $\rho_{\rm tomo}$ in Eq.~(\ref{eq:Tomography2P}), we have the analytic form,
\begin{eqnarray}
\mathcal{N}_{11}&=&\mathcal{N}_{22}=\kappa \int_0^{ \Delta T_w}  e^{-\kappa \tau} 
\left( (\cos \phi)^4 +(\sin \phi)^4  \right. \nonumber  \\
&&\left. +2(\sin\phi \cos \phi)^2  \cos(\Delta E_1 \tau )\right)  {\rm d}\tau ,  \\
\mathcal{N}_{12}&=&2 \kappa \int_0^{ \Delta T_w}  
(\sin\phi \cos \phi)^2 e^{-\kappa \tau} 
 \left( 1- \cos(\Delta E_1 \tau )\right)  {\rm d}\tau ,  \nonumber \\
\end{eqnarray}
where $\Delta E_1 \equiv E_{{\rm 1P},+}-E_{{\rm 1P},-}=\sqrt{(\Delta_1-\Delta_2)^2+4J^2}$.
Similarly, we find an expression for the off-diagonal element ($\mathcal{N}_{1122}=\mathcal{N}_{2211}^\ast$) as
\begin{eqnarray}
 \mathcal{N}_{1122}&=& 
 \kappa^2  \int_0^{t_f} {\rm d}t \int_0^{\Delta T_w} {\rm d} \tau 
 \langle  a_1^\dagger (t) a_1^\dagger (t+\tau) a_2 (t+\tau) a_2 (t) \rangle
\nonumber \\
&=&- \kappa  \int_0^{\Delta T_w}  e^{-\kappa \tau} 
\left(2(\sin\phi \cos \phi)^2   +(\cos \phi)^4  e^{-i \Delta E_1 \tau }  \right. \nonumber  \\
&&\left.+(\sin \phi)^4  e^{i \Delta E_1 \tau }\right)  {\rm d}\tau . 
\end{eqnarray}
From these results, the density matrix $\rho_{\rm tomo}$ obtained by the state tomography is found to be the same as those for pure $2002$ states, if the time window $\Delta T_w$ satisfies 
\begin{eqnarray}
 \Delta T_w \ll 1/\Delta E_1 . \label{WhichPathEraser}
\end{eqnarray}
Eq.~(\ref{WhichPathEraser}) can also be regarded as a condition for the ``which-path'' information of the emission frequencies to be erased~\cite{Scully, Troiani, Gies, Valle3}. With this short-time window, 
\begin{eqnarray}
\mathcal{N}_{11}=\mathcal{N}_{22}=-\mathcal{N}_{1122}=-\mathcal{N}_{2211}^\ast=\kappa \Delta T_w, 
\end{eqnarray}
and all other elements including $\mathcal{N}_{12}$ become zero to the leading order in $\Delta E_1 \Delta T_w$ ($\Delta E_1 \gg \kappa$ is already assumed). 
For the larger time window, $\Delta T_w >1/\Delta E_1 $, the photon correlation measurement does not reflect the initially prepared 2002-GES and the visibility of multi-photon interference degrades.
The rapid oscillation terms with a frequency $\Delta E_1$ in $\mathcal{N}_{ij}$ and $\mathcal{N}_{1122}$ arise from the coherent oscillation between $|G,10 \rangle$ and $|G,01 \rangle$, which takes place once the prepared $2002$-GES ($|E_s \rangle$) emits one photon until the second photon is emitted.

From the above consideration,  maximally available two-photon simultaneous detection rate $\mathcal{I}_{2002}(\equiv \mathcal{N}_{11} /\Delta T_{\rm rep})$ is estimated to be
\begin{eqnarray}
\mathcal{I}_{2002} =\frac{\kappa \Delta  T_w}{\Delta T_{\rm rep}} \le \frac{\kappa \Gamma _{2002}}{\Delta E_{1}} =\mathcal{O}\left( \frac{\kappa^2}{J}    (\sin{\varphi_s})^2  \right), \quad
\end{eqnarray}
for $ \Delta T_{\rm rep}(=t_f) \ge 1/\Gamma_{\rm 2002}$.
As an example, for parameters in Fig.~\ref{fig5}(c) $(t,\Delta_1, \Delta_2, \kappa)=(5,5,-0.05,0.1) \times \sqrt{2} g_{\rm 2P}$ and $g_{\rm 2P}=50 \ \mu {\rm eV}$ ($g=100$ $\mu$eV and $\chi=0.8$ meV), the maximum detection rate $\kappa \Gamma _{2002}/\Delta E_{1}$ is estimated to be 3.7 MHz.
This rate is by three orders faster than those obtained by SPDC-based source of kHz range~used in the reference~\cite{Nagata}. 

To obtain further enhancement of the rate $\mathcal{I}_{2002}$, the dynamic $Q$-switching in nanocavities~\cite{Tanaka} can be used, since high-$Q$ cavities are required only for the energy-selective pure-state excitation discussed in Sec.~\ref{2002exc} and not for the two-photon emission and detection processes studied in this section.
Deterministic and high-rate emission of the 2002-state photons will become possible by switching the $Q$ factor ($Q\propto 1/\kappa$) from the high value ($\kappa \ll J$) to low value ($\kappa \gg J$) just after the $\pi$-pulse preparation.
In this case, the maximally available rate is determined by the time scale of the two-photon Rabi dynamics, or the pulse duration $\delta t_p$ which is larger than $1/J$ (see the discussion in Sec.~\ref{2002exc}). 
The estimation gives an upper limit of the rate of $\mathcal{O}(J)$.

\section{\label{sec3} Extension to $N>2$}
Here we present extension of the above method for $2002$-state generation to general case of $N (>2)$.
Firstly, a general recipe for $N00N$-state generation is described in Sec.~\ref{Sec:Recipe}.
In the following section, Sec.~\ref{Sec:4004}, as example of the extension, we use the recipe to find design of four-photon $N00N$ state generator, where numerical simulation clarifies the requirement for system parameters (cavity $Q$ factor, detuning, coupling strength, etc.) to have pure $N00N$-state generation. Importance of step 2 in the recipe is stressed in Sec.~\ref{Sec:step2}. 
\subsection{\label{Sec:Recipe} Recipe for $N00N$-state generation }
As shown in the previous section for the case of $N=2$, the key of our method was to prepare the target $N00N$ state as an energy eigenstate. 
To do so, the quantum interference was utilized to eliminate the population in undesired states.
In similar manner, we can prepare $N00N$ state in the eigenstate of the system.
For clarity, we shall call such energy eigenstate, which can generate output of $N00N$-state photons, ``$N00N$-state generating eigenstate'' and abbreviate it as $N00N$-GES.
To prepare the $N00N$-GES for $N>2$, we just have to follow our recipe below consisting of three steps:
\begin{description}
 \item[(step 1)] \label{recipe1} Consider $N$-photon emitter coupling to two-mode cavities. 
Here, $N$-photon emitter is defined as quantum emitter which permits simultaneous emission of $N$ photons. 
System with $N_B$ QDs in biexciton state and $N_X$ QDs in a single exciton state is $N$-photon emitter of $N=2 N_B+N_X$. 
(System in Fig.~\ref{fig1} is two-photon emitter of $N=2$, $N_B=1$, and $N_X=0$.) 
Define the system Hamiltonian, according to the types of the coupling between emitter and cavity modes.
 \item[(step 2)] According to the Hamiltonian, draw schematic in the Hilbert subspace of $N_{\rm tot}$($\equiv$total number of photons and excitons)$=N$, which shows all directions of population flow at an initial time out from the prepared $N00N$-GES (Fig.~\ref{fig2}(a) is the corresponding schematic for the system in Fig.~\ref{fig1}).
The $N00N$-GES is a superposition state in the subspace of $N_{\rm tot}=N$ which does not contain $|G, n \ N-n \rangle$ for $1\le n \le N-1$, where $G$ represents a vacuum state of QDs with no exciton.
In the schematic, if there exist multiple or no path of the flow into each state to be eliminated ($|G, n \ N-n \rangle$ for $1\le n \le N-1$), the system can be considered as a candidate of $N00N$-state generator.
 \item[(step 3)] For the candidate system [(step 2)], solve $N$-conditioned eigenvalue problem to find tuning parameters of the system so that the $N00N$-GES be an eigenstate of the Hamiltonian.
\end{description}
We should add more explanation on step 2. 
In the schematic, directions of population flow are drawn by taking the $N00N$-GES as an initial state. 
It indicates the unitary dynamics of the density matrix at an initial time from the $N00N$-GES.
If the $N00N$-GES is the eigenstate of Hamiltonian, the density matrix should be conserved through the unitary dynamics.
To realize this situation, the population flow into the other state ($|G, n \ N-n \rangle$ with $1\le n \le N-1$) must be eliminated.
Presence of multiple or no path of the population flow into these unwanted states is a requirement so that the initially-prepared $N00N$-GES can be conserved in the unitary dynamics. If multiple flows into each of the other states are present, they need to interfere destructively to cancel out, for which additional requirement is taken into account in step 3. 
 
If the $N00N$-GES is contained as an eigenstate of the system, the resonant laser excitation can be used to generate it exclusively, being similar to the case of $N=2$.
In the following subsection, we apply this recipe to $4004$-state generation as an example.

\subsection{\label{Sec:4004}Example: 4004-state generation}
As an example, here we show that it is possible to generate $4004$ state using the recipe shown in the previous subsection. 

\begin{figure}[t] 
\begin{center}  
\includegraphics[scale=0.3]{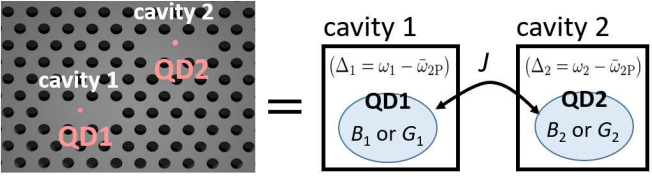} {}  
\end{center}
\vspace{-2 mm}
\caption{\label{fig7} Schematic of the QD-coupled nanocavity system (QD1 + QD2 + cavity 1 + cavity 2) for four-photon $N00N$ state generation (right), which can be realized in PhC platform (left).}
\end{figure}

\noindent
{\bf (step 1)---}We consider a system with two QDs, QD1 an QD2, in cavity 1 and cavity 2, respectively (Fig.~\ref{fig7}).
We assume that the interaction between the QDs and cavities occurs only through the biexciton-two photon transitions, in the same way as the 2002-state generator. 
Therefore, this system is categorized by $N=4$, $N_B=2$, and $N_X=0$ in recipe (i).
The biexciton two-photon resonance frequency in QD1 (QD2) is $\omega_{{\rm 2P},1(2)} \approx  \omega_{X_{1(2)}}-\chi_{1(2)}/2$.
State vectors are given by $| i_1 i_2 \rangle \otimes |n_1n_2 \rangle $, where $i_1 \in \{ B_1, G_1\}$ ($i_2\in \{ B_2, G_2\}$) represents the QD carrier states in for QD1 (QD2), and $ |n_1n_2 \rangle$ is a photon number state with $n_{1(2)}$ photons inside cavity 1(2). 
$B_j$ and $G_j$ ($j=1,2$) are biexciton and vacuum states in the $j$-th QD. 
Alternatively, we define more simple notation for the QD carrier states: $| Q \rangle \equiv |B_1, B_2 \rangle $, $| B1 \rangle \equiv |B_1, G_2 \rangle$, $| B2 \rangle \equiv |G_1, B_2 \rangle$, and $| G \rangle \equiv |G_1, G_2 \rangle$.
With this notation, total number of excitation is given by $N_{\rm tot} = 4 | Q \rangle \langle Q|+2( | B1 \rangle \langle B1|+ | B2 \rangle \langle B2|)+a_1^\dagger a_1+a^\dagger_2 a_2$.

The effective Hamiltonian in a frame rotating with the mean biexciton two-photon resonance frequency, $\bar{\omega}_{\rm 2P} (\equiv \frac{ \omega_{{\rm 2P},1}+\omega_{{\rm 2P},2} }{2}) $, $H_{\rm eff}=H -\bar{\omega}_{\rm 2P} N_{\rm tot}$, is given by
\begin{eqnarray}
H_{\rm eff} &=& \sum_{j=1,2} \Delta_{j} a^\dagger_j a_j +J(a^\dagger_1 a_2+a^\dagger_2 a_1) \nonumber \\
&& + \Delta_B \left(  |B1 \rangle \langle B1 | -  |B2 \rangle \langle B2 | \right)  \nonumber \\
&& +  \sum_{j=1,2}g_{ j} \left( a_j^2 |B_j \rangle \langle G_j | +   |G_j \rangle \langle B_j | (a_j^\dagger)^2 \right), \quad \label{Heff4}
\end{eqnarray}
where $\Delta_j (\equiv \omega_j -\bar{\omega}_{\rm 2P})$ is the cavity detuning, and $\Delta_B  (\equiv  \omega_{{\rm 2P},1}-\omega_{{\rm 2P},2})$ is the difference in the biexciton two-photon resonance frequencies.
Regarding $g_1$ as the unit of energy, there are five free parameters in the Hamiltonian which determine the system dynamics: ($g_2/g_1, J/g_1, \Delta_1/g_1, \Delta_2/g_1, \Delta_B/g_1$).

\begin{figure}[t]
\begin{center}  
\includegraphics[scale=0.30]{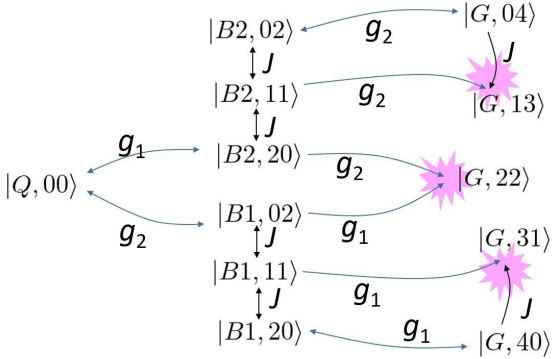} {}  
\end{center}
\vspace{-5 mm}
\caption{\label{fig8} Population flow at an initial time, shown in the Hilbert subspace of $N_{\rm tot}=4$ for the model system in Fig.~\ref{fig7}, starting from the prepared 4004-GES, i.e., Eq.~(\ref{4photon}) with Eqs.~(\ref{Req1:4004})-(\ref{Req4:4004}).}
\end{figure}

\noindent
{\bf (step 2)---}Based on the Hamiltonian $H_{\rm eff}$, we indicate the direction of population flow in the schematic Fig.~\ref{fig8}, in the Hilbert subspace $N_{\rm tot}=4$.
We found two paths of flow into $|G,13 \rangle $, $|G,22 \rangle $, and $|G,31 \rangle $, respectively.
Therefore, this system can be a candidate of $4004$-state generator.

\begin{figure}[t]
\begin{center}  
\includegraphics[scale=0.24]{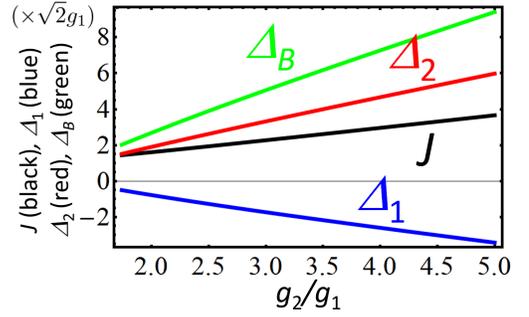} {}  
\end{center}
\vspace{-5 mm}
\caption{\label{fig9} Parameters with which the 4004-GES becomes one of the four-photon eigenstate $|E \rangle$ of $H_{\rm eff}$ for the model in Fig.~\ref{fig7}:
$J$ (black), $\Delta_1$ (blue), $\Delta_2$ (red), and $\Delta_B$ (green). }
\end{figure}

\noindent
{\bf (step 3)---}In this step, we will find the parameter values with which the 4004-GES is contained as one of the four photon eigenstate $|E \rangle$ of $H_{\rm eff}$, by solving the eigen problem $H_{\rm eff}|E \rangle =E|E \rangle $.
The four photon eigenstate is expanded by 12 states which appear in Fig.~\ref{fig8}:
\begin{eqnarray}
|E \rangle &\equiv & A_{Q,00}|Q,00 \rangle +\sum_{n1+n2=2} A_{B1,n_1n_2}|B1,n_1 n_2 \rangle  \nonumber \\
&&+\sum_{n1+n2=2}A_{B2,n_1n_2}|B2,n_1 n_2 \rangle \nonumber \\
&& +\sum_{n1+n2=4}A_{G,n_1n_2}|G,n_1 n_2 \rangle, \label{4photon}
\end{eqnarray}
for which we assign four requirements to be the 4004-GES, 
\begin{eqnarray}
&& A_{G,31}=0, \label{Req1:4004} \\
&& A_{G,22}=0, \label{Req2:4004} \\
&& A_{G,13}=0, \label{Req3:4004} \\
&& |A_{G,40}|=|A_{G,04}|. \label{Req4:4004}
\end{eqnarray}
Inserting the first three conditions, Eq.~(\ref{Req1:4004})-(\ref{Req3:4004}), into the eigen equation, they explicitly read
\begin{eqnarray}
0&=&\sqrt{4}J A_{G,04}+\sqrt{6}g_2 A_{B2,11}, \label{Req5:4004} \\
0&=&\sqrt{2}g_2 A_{B2,20}+\sqrt{2}g_1 A_{B1,02},  \label{Req6:4004}\\
0&=&\sqrt{6}g_1 A_{B1,11}+\sqrt{4}J A_{G,40}. \label{Req7:4004}
\end{eqnarray} 
These are understood as three requirements for these multiple quantum processes to vanish, in fully destructive way, the production rate of unwanted states ($|G,31 \rangle$, $|G,22 \rangle$, and $|G,13 \rangle$). 
The four requirements (Eqs.~(\ref{Req4:4004})-(\ref{Req7:4004})) fix the four free parameters  ($J/g_1, \Delta_1/g_1, \Delta_2/g_1, \Delta_B/g_1$) with $g_2/g_1$ remained as one free parameter.
Thanks to the symmetry of this system with respect to the exchange, (cavity 1, QD1)~$\leftrightarrow$~(cavity2, QD2), we can restrict our analysis, without loss of generality, to the reduced parameter space, $g_2/g_1 \ge 1$.
We found numerically the solutions to the conditioned eigenvalue equation for $g_2/g_1 \ge 1.74$, for which  the parameter values are shown in Fig.~\ref{fig9}. 

For parameters satisfying these requirements, the 4004-GES, $ | E_{4004} \rangle$, and the energy, $E_{4004}$, are given by
\begin{eqnarray}
&&|E_{4004}  \rangle = A_{4004} \left( \frac{|G,40 \rangle - |G,04 \rangle }{2} \right) +A_{Q,00}|Q,00 \rangle  \nonumber \\
&& \qquad + \sum_{l=B1,B2}\sum_{n_1+n_2=2} A_{l,n_1n_2}|l,n_1n_2 \rangle ,  \\
&& E_{4004} = \frac{6 \Delta_1+\Delta_B}{2} +\sqrt{\left(\Delta_1-\frac{\Delta_B}{2} \right)^2+6 g_1^2-4 J^2} \nonumber \\
 &&\quad =\frac{6 \Delta_2-\Delta_B}{2} -\sqrt{\left(\Delta_2+\frac{\Delta_B}{2} \right)^2+6 g_2^2-4 J^2} . \quad 
\end{eqnarray}
We notice that $|E_{4004}  \rangle$ contains also the photon number states with less than four photons, which, however, do not contribute to simultaneous four-photon detection and hence does not affect $\rho_{\rm tomo}$ reconstructed by four-photon quantum state tomography. 

\begin{figure}[t]
\begin{center}  
\includegraphics[scale=0.31]{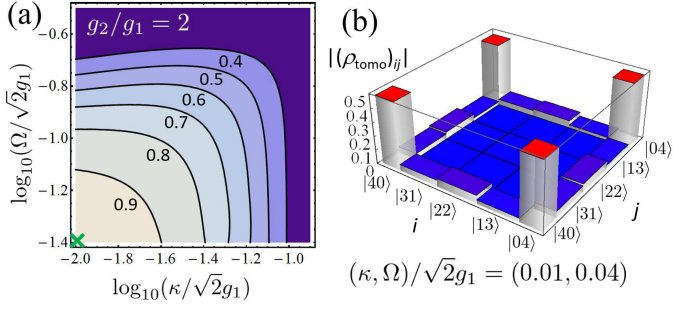} {}  
\end{center}
\vspace{-6 mm}
\caption{\label{fig10} (a) Simulated concurrence $\mathcal{C}$ of the emission from the model in Fig.~\ref{fig7}, shown as a function of ($\kappa, \Omega$) for $g_2/g_1=2$ and $(J, \Delta_1, \Delta_2, \Delta_B)= (1.61, -0.78, 1.90, 2.68) \times \sqrt{2}g_1$ (see Fig.~\ref{fig9}). (b) Four-photon density matrix $\rho_{\rm tomo}$, which will be constructed by quantum state tomography for $(\kappa, \Omega)=(0.01, 0.04)\times \sqrt{2}g_1$ (corresponding to a cross in (a)). }
\end{figure}

\noindent
{\bf Evaluation of purity and detection rate of the 4004-state emission---}
In order to evaluate the purity of emitted 4004-state photons, we study $\rho_{\rm tomo}$ and concurrence $\mathcal{C}(\equiv |\langle 40| \rho_{\rm tomo}|04 \rangle |)$ in similar manner as presented in Sec.~\ref{2002exc}. 
The simulation is performed for weak cw laser excitation on cavity 1 with $H_{\rm pump}=\Omega (a_1+a_1^\dagger)$, under the resonance condition, $\omega_p-\bar{\omega}_{\rm 2P}=E_{4004}/4$ (pumping on cavity 2 can also be used here, see appendix~\ref{appendix A}).
Fig.~\ref{fig10} (a) shows the simulated concurrence $\mathcal{C}$ plotted as a function of ($\kappa, \Omega$) for a set of parameters satisfying Eqs.~(\ref{Req4:4004})-(\ref{Req7:4004})$: 
g_2/g_1=2$ and $(J, \Delta_1, \Delta_2, \Delta_B)= (1.61, -0.78, 1.90, 2.68) \times \sqrt{2}g_1$ (see Fig.~\ref{fig9}).
Being similar to the case of $2002$-state (Fig.~\ref{fig5}), high concurrence, $\mathcal{C}>0.9$ is obtained only for high-$Q$ cavity (small $\kappa$) and weak pumping (small $\Omega$).
Fig.~\ref{fig10} (b) is the simulated $\rho_{\rm tomo}$ for $(\kappa, \Omega)=(0.01, 0.04)\times \sqrt{2}g_1$, which is close to that of a pure 4004 state.
However, demonstrating such a device with $\kappa=0.01\times \sqrt{2}g_1$ (i.e. $Q \approx 1.8 \times 10^6$ for $\omega_1 (\approx \omega_2)=1.3$ eV and $g_{\rm 1}=50 \ \mu$eV) is highly challenging with current state-of-the-art technology.

Compared to $2002$-state generation, much higher quality factor of cavities is necessary to have high $\mathcal{C}$. 
This indicates that the generation rate of pure $4004$ state is reduced more strongly than those of the $2002$ state. 
From an analogy with the discussion on detection rate of the $2002$ states, the maximally available rate of four-photon simultaneous detection is estimated to be $\propto \kappa \left(\kappa/J \right)^3$. 
However, this rate can be largely improved by using $Q$-swiching technique, in a similar manner as discussed for 2002-state generator in Sec.~\ref{2002rate}.

\subsection{\label{Sec:step2} Significance of step 2}
Step 2 of the recipe is useful as a general guideline to judge in a simple way (without any calculation) to which types of system configurations we can apply our scheme to design efficient $N00N$-state generator. 
As a simple example, we discuss the case of 4004 state generation.

\begin{figure}[t]
\begin{center}  
\includegraphics[scale=0.31]{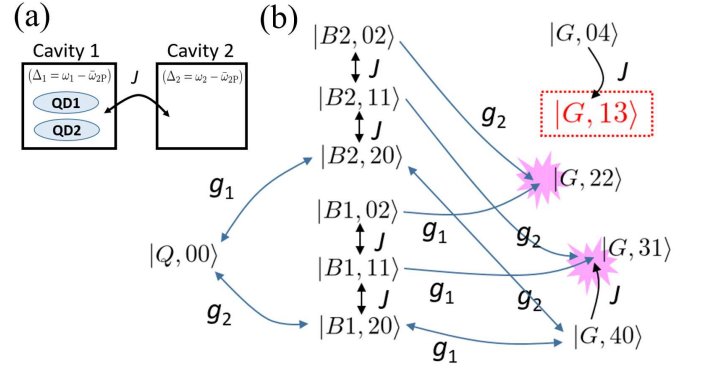} {}  
\end{center}
\vspace{-6 mm}
\caption{\label{fig11} (a) Two QDs in coupled-cavity system with a system configuration different from Fig.~\ref{fig8} (QDs emit photons to cavity 1 only through the biexciton two-photon transitions). 
(b) Coherent population flow at an initial time starting from prepared 4004-GES, i.e. Eq.~(\ref{4photon}) with Eqs.~(\ref{Req1:4004})-(\ref{Req4:4004}), for the system shown in (a). 
According to the recipe (step 2), the 4004-GES state cannot be included as an eigenstate in this system. Therefore, this system cannot be 4004-state generator with our method.}
\end{figure}

The system shown in Fig.~\ref{fig7} successfully becomes 4004-state emitter if system parameters are properly chosen.
With the same elements, two QDs and two cavities ($N=4$, $N_B=2$, and $N_X=0$), there is another system configuration as shown in Fig.~\ref{fig11} (a). The only difference from Fig.~\ref{fig7} is in the location of QDs.
However, by following step 2 of the recipe, we can see the system cannot be a candidate for $4004$-state generator with our method. 
To see this, we just have to draw schematic as shown in Fig.~\ref{fig11}(b), showing all population flows at an initial time out from the prepared 4004-GES, according to the Hamiltonian (in which the QD-cavity interaction term is replaced as $a_j^2 \to a_1^2$ and $(a_j^\dagger )^2 \to (a_1^\dagger)^2$ in Eq.~(\ref{Heff4})).
In the figure, we find multiple paths to $|G,22 \rangle $ and $|G,31 \rangle $ which are required to be eliminated for $4004$-state generation.
Full vanishing of the production rates for these two states is made possible by choosing parameters so that the multiple quantum paths interfere destructively.
This can be done by assigning two requirements,
\begin{eqnarray}
0&=& \sqrt{2} g_2 A_{B2,02}+ \sqrt{2} g_1 A_{B1,20}, \\
0&=& \sqrt{6}g_2 A_{B2,11}+\sqrt{6}g_1 A_{B1,11}+\sqrt{4} J A_{G,40},
\end{eqnarray}
to the four-photon eigen equation. 
On the other hand, there is only one population flow into $|G,13 \rangle$ (enclosed by dotted square in Fig.~\ref{fig11} (b)). 
Therefore, the destructive interference cannot be used, and the production rate does not vanish.
In this way, we can conclude the system in Fig.~\ref{fig11} (a) is not a candidate of 4004-state generator, irrespective of the parameters chosen.  
The situation is clearly different from the one shown in Fig.~\ref{fig8}.
Of course, this test (step 2) is cleared in the above-discussed $2002$-state generator, and also in polarization-entangled $2002$-state emitter~\cite{Gies}.

\section{\label{sec4} Conclusions}
In this paper, we proposed a generation method of photonic $N00N$ state with QDs in coupled nanocavities.
Starting from $N=2$, we show our recipe to generate $N00N$-state photons for $N>2$. 
The key of our method is to find the system parameters so that the $N00N$-state generating state ``$N00N$-GES'' can be prepared as an energy eigenstate of the system.
This is possible when multiple quantum paths can be used to eliminate perfectly the production rate of the unwanted states ($|G, n_1n_2 \rangle $ with $n_1+n_2=N$ and $1 \le n_1 \le N-1$).
In presence of the strong nonlinearity, the $N00N$-GES can be resonantly excited, and the $N00N$-state photons can be emitted and observed by simultaneous multi-photon detection.

To excite $N00N$-GES exclusively (to observe high concurrence $\mathcal{C}$ and small trace distance $\mathcal{D}$ from the pure state) through resonant pumping, the linewidth (or decay rate) of the $N00N$-GES needs to be small. 
Therefore, high-$Q$ cavities are required in this method. 
This limits the efficiency of multi-photon simultaneous emission and detection rates of the $N00N$-state photons, $\mathcal{I}_{N00N}$, which follows scaling law, $\mathcal{I}_{N00N} \propto \kappa (\kappa/J)^{N-1}$. 
Even if we take into account the limitation, the available detection rate in our $2002$-state generator is estimated to be by three orders higher than those obtained with the typical $2002$-state photon source~\cite{Nagata}. 
Moreover, by utilizing the $Q$-switching technique~\cite{Tanaka}, further enhancement in the emission rate will become possible (the emission can even become deterministic for the $2002$-state generator). 

We mention here the dephasing effect of the QD excitons (say $\gamma_{\rm phase}$), which was simply neglected in the analysis.
The most dominant effect will come from that for the biexciton state (not for the single-exciton state) for the $N00N$-state generator using the biexciton-two-photon resonance.
Given that the $\pi$-pulse excitation of $2002$-GES is successfully achieved, the population decay dynamics and the two-photon detection rate (in Sec.~\ref{2002decay} and Sec.~\ref{2002rate}) are unaffected by the presence of the dephasing.
The reason for the latter is that the fast oscillation terms in the $\tau$-dependency e.g. in Eq.~(\ref{eq:Nij}), attribute solely to the dynamics of the one-photon states with no QD exciton. 
As for the $\pi$-pulse excitation, the Rabi excitation dynamics can be made fast enough to be completed before the QD biexciton can dephase, if $\gamma_{\rm phase}<\mathcal{O}(J)=\mathcal{O}(g_{\rm 2P})$. 
Considering the rough estimation, $g_{\rm 2P}=50 \ \mu$eV, the last requirement is fulfilled and thus we can use our scheme robustly, as far as the dephasing is not too strong. Of course, we cannot apply the discussion directly to the general case of $N00N$-state generator with $N>2$. In this case, we can set the detection time window as $\Delta T_w < 1/\gamma_{\rm phase}$ so that the dephasing effect is reduced in the multi-photon detection. 

We also mention the difference between our method and the previously-reported method for polarization-entangled two-photon $N00N$ state~\cite{Gies,Valle3}. 
As discussed above, key of our method is to prepare $N00N$-GES as an energy eigenstate, while the previous method relies on the spontaneous emission of the prepared biexciton state into degenerate two polarization modes.
Therefore, the previous method requires high symmetry between different polarization modes, i.e. pure $2002$ state is generated when two cavity modes have the same strength in the biexciton-two-photon coupling and same frequency at the biexciton-two-photon resonance.
On the other hand, with our method, it is possible to generate pure $2002$-state photons even if their coupling strength are different, as far as the requirement in Eq.~(\ref{Cond1}) is fulfilled.   

Our method also has a disadvantage.
$N00N$-state generator proposed here requires high-$Q$ cavity, strong coupling $g$, and high-precision tuning of cavity resonance.
The realization becomes harder and harder as $N$ increases.
Further optimization of the system design, including the parameter choice in combination with frequency filtering~\cite{Valle3, Kamide2}, will increase the quality of the emitted $N00N$-state photons, and hence could relax the requirements, of which details we leave as a future issue.

\acknowledgements
We thank T. Horikiri, M. Yamaguchi, M. Bamba, and M. Holmes for useful comments and discussions.
This work was supported by the JSPS KAKENHI (15H05700, 15K20931), the Project for Developing Innovation Systems of MEXT, and New Energy and Industrial Technology Development Organization (NEDO).
 
\appendix

\section{\label{appendix A} Selection rule in resonant two-photon excitation specific to the 2002-state generator}
In Sec.~\ref{2002exc}, we see that the 2002-GES, $|E_s \rangle$, can be excited by resonant laser field applied to cavity 2, $H_{{\rm pump},2}=\Omega (a_2+a_2^\dagger)$, but not to cavity 1.
Actually, we confirmed by numerical simulation that resonant excitation on cavity 1, by replacing the pump Hamiltonian with $H_{\rm pump}=\Omega (a_1+a_1^\dagger)$, cannot generate the target 2002-state photons.
Here, we show that it is due to an underlying selection rule for resonant two-photon excitation, which is specific to this 2002 generator. (As for the 4004 generator in Sec.~\ref{Sec:4004}, we confirmed numerically that the target 4004-state photons can be generated for each case with resonant laser field on cavity 1 or cavity 2.)
This is an interesting physics (possibly with some application), which although is out of the main scope of this paper.

In order to see this, we apply second-order perturbation theory based on Schrieffer-Wolff transformation~\cite{Schrieffer, Valle1} to find the effective Hamiltonian, $\tilde{\mathcal H}=e^{\mathcal S} {\mathcal H}e^{\mathcal -S}$, and the two-photon transition matrix element, $\langle E_s |\tilde{\mathcal H}| G,00\rangle$. 
To construct the effective Hamiltonian, we focus on the Hilbert subspace spanned by four eigenstates, the vacuum state, $|G,00 \rangle$, 2002-GES, $|E_s \rangle$, and two one-photon states, $|{\rm 1P},\pm \rangle$.
The other three two-photon eigenstates are energetically separated from the 2002-GES, hence are not considered here (e.g., in Fig.~\ref{fig5} (a), two of them are shown at $N_{\rm tot}=2$ and another is outside the plot range).

Firstly, we divide Hamiltonian (in the rotating frame with the excitation laser frequency) into two terms, $\mathcal{H}=H_0+V$: the unperturbed term, 
\begin{eqnarray} 
H_0&=&0 \cdot |G,00 \rangle \langle G,00|+(E_s-2(\omega_p-\omega_{\rm 2P})) |E_s \rangle \langle E_s| \nonumber \\
&+& \sum_{\sigma=\pm}(E_{{\rm 1P},\sigma}-(\omega_p-\omega_{\rm 2P})) |{\rm 1P},\sigma \rangle \langle {\rm 1P},\sigma |,
\end{eqnarray}
and perturbation term, $V=V_1$ or $V_2$ with $V_{1(2)}=\Omega (a_{1(2)} +a_{1(2)}^\dagger)$.
Using Schrieffer-Wolff transformation~\cite{Schrieffer, Valle1}, the general form of the effective Hamiltonian upto the second order in $V$ is
\begin{eqnarray}
\tilde{\mathcal H} &\approx& H_0 +\sum_{i,f} M_{fi} |f \rangle \langle i|,\\
M_{fi}&=& \sum_k \frac{ \langle f| V |k \rangle \langle k | V|i \rangle}{2} \left(\frac{1}{E_f-E_k}+\frac{1}{E_i-E_k} \right), \nonumber  \\
&& \label{2002matrix}
\end{eqnarray}
where $H_0|m \rangle =E_{m} |m \rangle $ for $m=i,f,k$.
The matrix element for the two-photon transition to the target 2002-GES under the resonant excitation ($E_s-2(\omega_p-\omega_{\rm 2P})=0$) is obtained by putting the initial state $|i \rangle =|G,00 \rangle$, final target state $|f \rangle =|E_s \rangle$, and two intermediate states $|k \rangle = |{\rm 1P},\sigma(=\pm) \rangle$  into Eq.~(\ref{2002matrix}).

Following straight forward calculation, we found the two-photon transition matrix element, for resonant pumping on cavity 2 ($V=V_2$), 
\begin{eqnarray}
M_{fi}&=&\sin{\varphi_s} \times \Omega^2 \nonumber \\
& \times & \left(
\frac{(\sin{\phi})^2}
{E_{{\rm 1P},+}-(\omega_p-\omega_{\rm 2P})}
+ \frac{(\cos{\phi})^2}
{E_{{\rm 1P},-}-(\omega_p-\omega_{\rm 2P})}
 \right), \nonumber \\
&& \label{Mfi_cav2}
\end{eqnarray}
while for resonant pumping on cavity 1 ($V=V_1$), 
\begin{eqnarray}
M_{fi}&=&-\sin{\varphi_s} \times \Omega^2  \nonumber \\
& \times & \left(
\frac{(\cos{\phi})^2}
{E_{{\rm 1P},+}-(\omega_p-\omega_{\rm 2P})}
+ \frac{(\sin{\phi})^2}
{E_{{\rm 1P},-}-(\omega_p-\omega_{\rm 2P})}
 \right). \nonumber \\
&& \label{Mfi_cav1}
\end{eqnarray}
The two terms in the bracket in Eq.~(\ref{Mfi_cav2}) and Eq.~(\ref{Mfi_cav1}) correspond to the contribution from two transition paths via two intermediate states $|{\rm 1P}, \pm \rangle$.  
Using the condition for cavity and laser frequencies, $\omega_p-\omega_{\rm 2P}=\Delta_2=E_s/2$, and definition for $\phi$ below Eq.~(\ref{1Pstate2}) (see Sec.~\ref{2002design} and Sec.~\ref{2002exc}), we found the two contribution perfectly cancels with each other in Eq.~(\ref{Mfi_cav1}), leading to $M_{fi}=0$ for the pumping on cavity 1 ($V=V_1$).
On the other hand, this is not the case, $M_{fi}\ne 0$, for the pumping on cavity 2 ($V=V_2$). 
The result clearly shows existence of a selection rule for the two-photon resonant excitation:
excitation of the $2002$-GES through cavity pumping is allowed only via cavity 2, but forbidden via cavity 1.

\section{\label{appendix B} Method of simulation for state tomography, trace distance, and concurrence}
Here, the method of numerical calculation is briefly summarized.  
The two-photon density matrix $\rho_{\rm tomo}$, which can be experimentally reconstructed by the quantum state tomography~\cite{Kwiat, Troiani, Gies, Valle3}, is obtained from the two-photon correlation functions; 
\begin{eqnarray}
\rho_{\rm tomo} =\mathcal{Z}
\left(
    \begin{array}{ccc}
\langle \frac{a_1^\dagger a_1^\dagger a_1 a_1}{2}\rangle  & \langle \frac{a_1^\dagger a_1^\dagger a_2 a_1}{\sqrt{2}} \rangle &  \langle \frac{a_1^\dagger a_1^\dagger a_2 a_2}{2} \rangle \\
\langle \frac{a_1^\dagger a_2^\dagger a_1 a_1}{\sqrt{2}} \rangle  & \langle a_1^\dagger a_2^\dagger a_2 a_1 \rangle &  \langle \frac{a_1^\dagger a_2^\dagger a_2 a_2}{\sqrt{2}} \rangle  \\
    \langle \frac{a_2^\dagger a_2^\dagger a_1 a_1}{2} \rangle &   \langle \frac{a_2^\dagger a_2^\dagger a_2 a_1}{\sqrt{2}} \rangle &\langle \frac{a_2^\dagger a_2^\dagger a_2 a_2}{2} \rangle
    \end{array}
  \right),  \ 
\label{eq:Tomography2P} 
\end{eqnarray}
where $\langle X \rangle \equiv {\rm Tr}(X \rho)$ and $\mathcal{Z}$ is a normalization factor to give ${\rm Tr}(\rho_{\rm tomo})=1$.
Each component of the matrix is measured by simultaneous two-photon countings.
Concurrence, $\mathcal{C} (\equiv 2|(\rho_{\rm tomo})_{1,3}|)$, is an entanglement measure, given by the off-diagonal matrix element of $\rho_{\rm tomo}$. 
This measure becomes unity for pure $2002$ states~\cite{Wootters},
\begin{eqnarray}
\rho_{2002,\theta}=\frac{(|20\rangle +e^{-i\theta}|02\rangle)(\langle 20|+e^{i\theta}\langle 02|)}{2}.
\end{eqnarray}
The trace distance, $\mathcal D$, is defined by $\mathcal D \equiv {\rm Tr}|\rho_{2002,\theta}- \rho_{\rm tomo}| = {\rm Tr}\sqrt{(\rho_{2002,\theta}- \rho_{\rm tomo})^\dagger (\rho_{2002,\theta}- \rho_{\rm tomo}) }$. 
This is a measure indicating how close to a pure 2002 state the observed photons are, where $\mathcal D=0$ means that the system emit exactly the pure 2002-state photons. 
In the simulation, the phase $\theta (\approx \pi)$ is chosen to minimize $\mathcal D$.

\section{\label{appendix C} Two-photon emission rate of the 2002-state generator under weak cw laser pumping}
By using the result in appendix~\ref{appendix A}, we obtain an approximate analytic expression for the two-photon emission rate of the 2002-state generator in the case of weak cw excitation (via cavity 2).
The effective Hamiltonian is 
\begin{eqnarray} 
\tilde{\mathcal H}&\approx &\xi ( |G,00 \rangle \langle E_s|+ |E_s \rangle \langle G,00|) \nonumber \\
&+& \sum_{\sigma=\pm}(E_{{\rm 1P},\sigma}-(\omega_p-\omega_{\rm 2P})) |{\rm 1P},\sigma \rangle \langle {\rm 1P},\sigma |,
\end{eqnarray}
where we used the above result for the two-photon transition amplitude, $\xi$ ($\equiv M_{fi}$ in Eq.~(\ref{Mfi_cav2}) in appendix~\ref{appendix A}), and neglected slight energy shift in the diagonal part of $\tilde{\mathcal H}$ for simplicity. 
Using this simple Hamiltonian, we found steady state solution to the quantum Master equation, $\dot{\rho}=-i[\tilde{\mathcal H}, \rho]+\mathcal{L}_{\rm loss} \rho =0$, in analytic form (see the main text for the definition of $\mathcal{L}_{\rm loss}$).
For the steady state, excited population of the $2002$-GES is 
\begin{eqnarray}
P_{2002}=4 \xi^2/\Gamma_{2002}^2
\end{eqnarray} 
to the leading order in $\xi$, where $\Gamma_{2002}=2(\sin{\varphi_s})^2 \kappa$ is the decay rate. 
The result is directly related to the rate of simultaneous two-photon emission (detected within a time window $\Delta T_w$ of relative delay) from the same cavity,
\begin{eqnarray}
\mathcal{I}_{2002}&=&\kappa^2 \Delta T_w  \langle  {a_1^\dagger}^2 a_1^2 + {a_2^\dagger}^2 a_2^2 \rangle  \nonumber \\
&=& 2 \kappa^2 \Delta T_w (\sin{\varphi_s})^2 P_{2002}\nonumber \\
 &=& 8 \kappa^2 \Delta T_w (\sin{\varphi_s})^2 \xi^2/\Gamma_{2002}^2.
\end{eqnarray}
where the time window is assumed small for the detector.
The analytic result explains the high contrast in the peak emission intensity found at the two conditions, $\Delta_2=E_{\pm}/2$, in Fig.~\ref{fig3} (dashed line).

\end{document}